\documentclass[preprint,showpacs,preprintnumbers,epsbox]{revtex4}
\usepackage{graphicx}
\usepackage{dcolumn}
\usepackage{bm}

\def\lesssim{\ \raise.3ex\hbox{$<$}\kern-0.8em\lower.7ex\hbox{$\sim$}\ }
\def\gesim{\ \raise.3ex\hbox{$>$}\kern-0.8em\lower.7ex\hbox{$\sim$}\ }
\font\scripti=cmmi7
\font\scriptscripti=cmmi5
\def\sib#1{\setbox0 = \hbox{\scripti #1}
  \kern-.02em\copy0\kern-\wd0
  \kern.04em\box0} 
\def\ssib#1{\setbox0 = \hbox{\scriptscripti #1}
  \kern-.02em\copy0\kern-\wd0
  \kern.04em\box0} 
\font\tenib=cmmib10 
\skewchar\tenib='177 \skewchar\tenib='177 \skewchar\tenib='177
\def\pbold#1{\setbox0 = \hbox{$ #1 $}
  \kern-.022em\copy0\kern-\wd0
  \kern.011em\copy0\kern-\wd0
  \kern.011em\copy0\kern-\wd0
  \kern.011em\copy0\kern-\wd0
  \kern.011em\box0} 
\begin{document}
\title{Effective interaction between molecules in the BEC regime of a superfluid Fermi gas}
\author{Y. Ohashi}
\affiliation{Institute of Physics, University of Tsukuba, Tsukuba,
  Ibaraki 305, Japan,\\  Department of Physics, University of Toronto, Toronto,
  Ontario, Canada M5S 1A7}
\date{\today}
\begin{abstract}
We investigate the effective interaction between Cooper-pair molecules in the strong-coupling BEC regime of a superfluid Fermi gas with a Feshbach resonance. 
Our work uses a path integral formulation and a renormalization group (RG) analysis of fluctuations in a single-channel model. We show that a physical cutoff energy $\omega_c$ originating from the finite molecular binding energy is the key to understanding the interaction between molecules in the BEC regime. Our work thus clarifies recent results by showing that $a_{\rm M}=2a_{\rm F}$ is a {\it bare} molecular scattering length while $a_{\rm M}=(0.6\sim0.75) a_{\rm F}$ is the low energy molecular scattering length renormalized to include high-energy scattering up to $\omega_c$ (here $a_{\rm F}$ is the scattering length between Fermi atoms). We also include many-body effects at finite temperatures. 
We find that $a_{\rm M}$ is strongly dependent on temperature, vanishing at $T_{\rm c}$, consistent with the earlier Bose gas results of Bijlsma and Stoof.
\end{abstract}
\maketitle
%
\par
Recently, the interaction between bound molecules has attracted much attention in the strong-coupling BEC regime of a superfluid Fermi gas\cite{Randeria,Engelbrecht,Strinati,Petrov}. In a trapped Fermi gas with a Feshbach resonance, one can tune the magnitude of a pairing interaction by varying the threshold energy of the Feshbach resonance\cite{Timmermans,Holland,Ohashi}. Using this tunable interaction, the BCS-BEC crossover\cite{Ohashi} has been observed in $^{40}$K and $^6$Li, where the character of superfluidity continuously changes from the weak-coupling BCS type to the BEC of tightly bound molecules, as one increases the pairing interaction\cite{Randeria,Nozieres}. In the strong-coupling BEC regime, where the molecules already form above the superfluid phase transition temperature $T_{\rm c}$\cite{Ohashi}, the system can be regarded as a molecular Bose gas. 
\par
The magnitude of the molecular interaction in the BEC regime was originally studied in the superconductivity literature\cite{Randeria,Engelbrecht}, where the result $a_{\rm M}=2a_{\rm F}$ was obtained using a Ginzburg-Landau (GL) type expansion. Here $a_{\rm M}$ is the molecular scattering length and $a_{\rm F}$ is the atomic $s$-wave two-body scattering length. Later, Pieri and Strinati\cite{Strinati} obtained $a_{\rm M}\simeq 0.75a_{\rm F}$ within a two-body $t$-matrix approximation. More recently, Petrov and co-workers\cite{Petrov} obtained $a_{\rm M}\simeq 0.6a_{\rm F}$ by solving a four-fermion problem. This last result is confirmed by a direct Monte-Carlo simulation for the ground state\cite{Giorgini}. In addition, Grimm and co-workers also obtained $a_{\rm M}\simeq 0.6a_{\rm F}$ in superfluid $^6$Li\cite{Grimm}, using a Thomas-Fermi fit to their molecular condensate density profile.
\par
In this paper, we investigate this interaction between molecules in the strong-coupling BEC regime of a uniform Fermi gas, using a single-channel model. 
We show that the finite binding energy of a Cooper-pair ``molecule'' naturally leads to a {\it physical} cutoff energy $\omega_c$, which is not present in the usual {\it atomic} BEC. In a two-molecule case, we show that $\omega_c$ is the key to understanding the physics behind the difference between earlier results, namely, $a_{\rm M}=2a_{\rm F}$\cite{Randeria,Engelbrecht} and $a_{\rm M}=(0.6\sim0.75)a_{\rm F}$\cite{Strinati,Petrov}. 
We also calculate the many-body $t$-matrix at finite temperatures, extending the renormalization group analysis\cite{Stoof} for an atomic BEC to the BEC regime of a superfluid Fermi gas. 
In a many-particle system, we show that $a_{\rm M}$ depends strongly on temperature, vanishing in the region near $T_{\rm c}$. 
\par
We consider a two-component Fermi gas described by the single-channel BCS model. In the functional integral formalism, the action $S$ in the partition function $Z\equiv \sum_\sigma \int {\cal D}\Psi^\dagger_\sigma{\cal D}\Psi_\sigma e^{-S}$ has the form\cite{Engelbrecht},
$S
=\int_0^\beta d\tau\int d{\bf r}
\Bigl[
\sum_\sigma
\Psi_\sigma^\dagger
\Bigl(
{\partial \over \partial \tau}+{{\hat {\bf p}}^2 \over 2m}-\mu
\Bigr)
\Psi_\sigma
-
U\Psi^\dagger_\uparrow\Psi^\dagger_\downarrow\Psi_\downarrow\Psi_\uparrow
\Bigr].$
Here, $\Psi_\sigma({\bf r},\tau)$ and $\Psi_\sigma^\dagger({\bf r},\tau)$ are a Grassmann variable and its conjugate, describing Fermi atoms with pseudo-spin $\sigma=\uparrow,\downarrow$. $\mu$ is the atomic chemical potential. In $^{40}$K and $^6$Li Fermi gases, the tunable pairing interaction $U$ is associated with the Feshbach resonance\cite{Ohashi,Timmermans,Holland}. We simply treat $U$ as a tunable parameter in a single channel formulation. This is valid for a broad Feshbach resonance. 
\par
To discuss the strong-coupling BEC regime, it is convenient to introduce a Cooper-pair Bose field $\Delta({\bf r},\tau)$, using the usual Stratonovich-Hubbard transformation\cite{Engelbrecht}. After functional integrations over $\Psi_\sigma$ and $\Psi_\sigma^\dagger$, we obtain $Z=\int {\cal D}\Delta^\dagger{\cal D}\Delta e^{-S_\Delta}$, where
$S_\Delta=
\int_0^\beta d\tau\int d{\bf r}
{|\Delta|^2 \over U}
-{\rm Tr}
\Bigl[
\ln[-{\hat G}^{-1}]
\Bigr]$\cite{Engelbrecht}. The fermion single-particle Green's function is given by
\begin{eqnarray}
{\hat G}^{-1}
\equiv-{\partial \over \partial\tau}-({{\hat {\bf p}}^2 \over 2m}-\mu)\tau_3
+
\left(
\begin{array}{cc}
0 & \Delta({\bf r},\tau)\\
\Delta^\dagger({\bf r},\tau)
&
0 
\end{array}
\right),
\end{eqnarray}
where $\tau_j$ ($j=1,2,3$) is the Pauli matrices. The mean-field gap equation is obtained from the saddle point solution for $\Delta_{\rm MF}$ determined by $\partial S_\Delta/\partial\Delta=0$. Expanding the action $S_\Delta$ around this mean-field solution ($\Delta_{\rm MF}$), we obtain
\begin{eqnarray}
S_\Delta
&=&
\int_0^\beta d\tau\int d{\bf r}
{|\delta\Delta|^2 \over U}
\nonumber
\\
&+&\sum_{n=2}^\infty
{(-1)^n \over n}
{\rm Tr}
\Biggl[
\Bigl[
{\hat G}^0
\left(
\begin{array}{cc}
0 & \delta\Delta \\
\delta\Delta^\dagger& 0
\end{array}
\right)
\Bigr]^n
\Biggr],
\label{eq.4}
\end{eqnarray}
where $\delta\Delta({\bf r},\tau)\equiv\Delta({\bf r},\tau)-\Delta_{\rm MF}$ describes fluctuations in the particle-particle channel. The mean-field $2\times2$ matrix Green's function ${\hat G}^0$ has the standard BCS form,
\begin{eqnarray}
{\hat G}^0({\bf p},i\omega_m)
=-
{i\omega_m+(\varepsilon_{\bf p}-\mu)\tau_3-\Delta_{\rm MF}\tau_1 
\over 
\omega_m^2+E_{\bf p}^2}.
\label{eq.5}
\end{eqnarray}
Here, $i\omega_m$ is the fermion Matsubara frequency. $E_{\bf p}\equiv\sqrt{(\varepsilon_{\bf p}-\mu)^2+\Delta_{\rm MF}^2}$ is the single-particle excitation spectrum, where $\varepsilon_{\bf p}$ is the kinetic energy of a Fermi atom. 
\par
In the BEC regime, the atomic chemical potential $\mu$ is related to the atomic scattering length $a_{\rm F}$ by the formula ${\bar \mu}\equiv-1/2ma_{\rm F}^2$\cite{Randeria,Engelbrecht}. We note that ${\bar \mu}$ is large and negative as we approach the BEC limit ($a_{\rm F}\to+0$). In this regime, since the BCS order parameter at $T=0$ is given by $\Delta_{\rm MF}=\sqrt{16/3\pi}|{\bar \mu}|^{1/4}\varepsilon_{\rm F}^{3/4}$\cite{Engelbrecht}, we find $|{\bar \mu}|\gg\Delta_{\rm MF}$. Thus, we can use approximation $E_{\bf p}\simeq\varepsilon_{\bf p}+|{\bar \mu}|$ in (\ref{eq.5}). In addition, the energy gap $E_g$ is given by $E_g=\sqrt{|\mu|^2+\Delta_{\rm MF}^2}\simeq |{\bar \mu}|$, so that the binding energy $E_{\rm bind}$ of a Cooper-pair is given by $E_{\rm bind}=2E_g=2|{\bar \mu}|$.
\par
In the BEC regime, we expand (\ref{eq.4}) in powers of $\delta\Delta$, retaining terms up to $n=4$. We recall that the Gaussian approximation\cite{Randeria,Engelbrecht,Ohashi,Nozieres} only keeps fluctuation terms to order $n=2$. 
In the superfluid phase, we find
\begin{eqnarray}
&S_\Delta&
=
{1 \over \beta}
\sum_{q}
\Phi_q^\dagger
(-i\nu_n+\varepsilon_q^M+U_Mn_c)
\Phi_q
\nonumber
\\
&+&
{n_cU_M \over 2}
{1 \over \beta}
\sum_{q}
(\Phi_q\Phi_{-q}+\Phi_{-q}^\dagger\Phi_q^\dagger)
\Bigr]
\nonumber
\\
&+&{1 \over \beta^2}
\sum_{q,K}V_{\rm eff}^{(3)}(q,K)
(\Phi_{{K \over 2}+q}^\dagger\Phi_{{K \over 2}-q}^\dagger\Phi_K+\Phi_K^\dagger\Phi_{{K \over 2}-q}\Phi_{{K \over 2}+q})
\nonumber
\\
&+&
{1 \over 2\beta^3}
\sum_{q,q',K}V_{\rm eff}^{(4)}(q,q',K)
\Phi_{{K \over 2}+q'}^\dagger\Phi_{{K \over 2}-q'}^\dagger\Phi_{{K \over 2}-q}\Phi_{{K \over 2}+q}.
\label{eq.7}
\end{eqnarray}
Here, $U_M\equiv 4\pi(2a_{\rm F})/M$ (where $M=2m$) is the bare $s$-wave molecular interaction obtained in what amounts to the time-dependent GL theory\cite{Randeria,Engelbrecht}; $\varepsilon_{\bf q}^M\equiv q^2/2M$ is the kinetic energy of a free molecule, and $i\nu_n$ is the pair boson Matsubara frequency, with $p\equiv ({\bf p},i\omega_m)$. 
In the BEC limit of interest, the fluctuations in the order parameter $\delta\Delta_q$ 
involve the molecular condensate. We introduce a renormalized molecular field $\Phi_q\equiv \delta\Delta_q/\eta$, where $\eta=[\sum_{\bf p}1/4(\varepsilon_{\bf p}-{\bar \mu})^2]^{-1/2}$. 
Defining $\phi_M\equiv \Delta_{\rm MF}/\eta$ as the equilibrium value of the Bose-condensate order parameter, at $T=0$, one can show that $n_c\equiv \phi_M^2=n_{\rm F}/2$ (where $n_{\rm F}$ is the number of Fermi atoms) gives the number of Bose-condensed molecules. 
\par
\begin{figure}
\includegraphics[width=9cm,height=6cm]{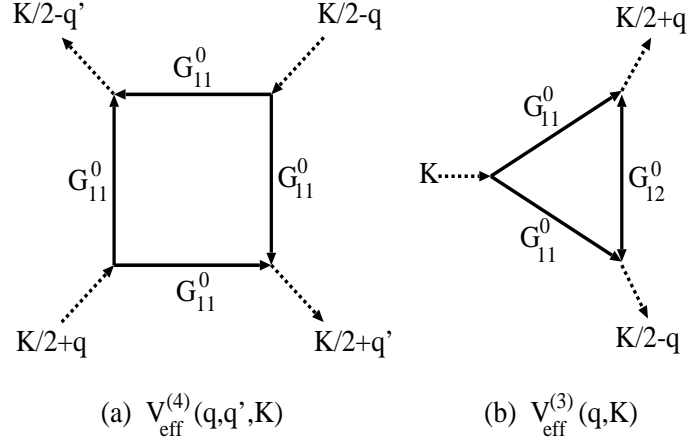}%
\caption{
(a) Interaction between molecules $V_{\rm eff}^{(4)}$ mediated by Fermi atoms. The dotted lines describe the molecular Boson propagator. (b) Particle-non-conserving interaction $V_{\rm eff}^{(3)}$.
\label{fig1}
}
\end{figure}

\par
$V_{\rm eff}^{(4)}$ in (\ref{eq.7}) describes an interaction between molecules mediated by the Fermi gas [see Fig. 1(a)], given by
\begin{eqnarray}
V_{\rm eff}^{(4)}(q,q',K)
&=&
{\eta^4 \over \beta}
\sum_p
G^0_{11}(p+{q-q' \over 2}+{K \over 4})
G^0_{11}(-p+{q+q' \over 2}+{K \over 4})
\nonumber
\\
&\times&
G^0_{11}(-p-{q+q' \over 2}+{K \over 4})
G^0_{11}(p-{q-q' \over 2}+{K \over 4}).
\label{eq.8}
\end{eqnarray}
Here, $G_{11}^{0}$ is the diagonal (1,1)-component of the BCS mean-field Green's function defined in (\ref{eq.5}). $V_{\rm eff}^{(3)}$ in (\ref{eq.7}) represents a particle-non-conserving interaction, given by [see also Fig. 1(b)]
\begin{eqnarray}
V_{\rm eff}^{(3)}(q,K)
=
-{\eta^3 \over \beta}
\sum_p
G^0_{11}(p+{K \over 2})G^0_{11}(-p+{K \over 2})G^0_{12}(p+q).
\label{eq.9}
\end{eqnarray}
This interaction involves the off-diagonal (anomalous) Green's function $G_{12}^{0}$ defined in (\ref{eq.5}), so that $V_{\rm eff}^{(3)}$ only exists in the superfluid phase. 
\par
\begin{figure}
\includegraphics[width=9cm,height=6cm]{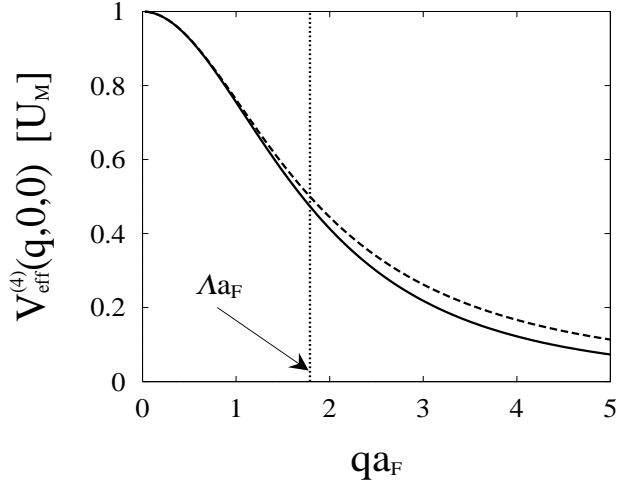}%
\caption{
\label{fig2}
Calculated momentum dependence of the boson-boson interaction $V_{\rm eff}^{(4)}(q,0,0)$ (solid line). The dashed line is the Lorentzian fit given by $U_{\rm M}/(1+(q/\Lambda)^2)$. The vertical dotted line shows the position of the cutoff momentum $\Lambda$.
}
\end{figure}
\par
For $q=q'=K=0$, one can show that $V_{\rm eff}^{(4)}(0,0,0)=U_M$ and $V_{\rm eff}^{(3)}(0,0)=\sqrt{n_c}U_M$, where $U_M=4\pi(2a_{\rm F})/M$. 
Using these results in place of $V_{\rm eff}^{(4)}(q,q',K)$ and $V_{\rm eff}^{(3)}(q,K)$ in (\ref{eq.7}), we reproduce the previous results obtained in the literature, namely, $a_{\rm M}=2a_{\rm F}$\cite{Randeria,Engelbrecht}. However, $V_{\rm eff}^{(4)}$ and $V_{\rm eff}^{(3)}$ actually depend on momentum, and go to zero with a characteristic cutoff momentum $\Lambda$. 
To calculate this cutoff momentum $\Lambda$, we expand (\ref{eq.8}) and (\ref{eq.9}) in terms of the relative momentum (taking the center-of-mass-momentum ${\bf K}$ to be zero\cite{notez}), keeping terms of $O(q^2)$. 
We obtain $V_{\rm eff}^{(4)}(q,q',0)\simeq U_M F(q)F(q')$ and $V_{\rm eff}^{(3)}(q,0)\simeq \sqrt{n_c}U_M F(q)$, where $F(q)=1-(q/\Lambda)^2$. The characteristic momentum is found to be $\Lambda= 4/\sqrt{5}a_{\rm F}$. For larger values of the momentum, we replace $F(q)$ by the Lorentzian $[1+(q/\Lambda)^2]^{-1}$. This gives a very good fit to the momentum dependent interaction, as shown in Fig. 2. In terms of energy, we can write $F(q)=[1+\varepsilon^M_{\bf q}/\omega_c]^{-1}$, where the interaction or molecular cutoff energy is defined by
\begin{eqnarray}
\omega_c\equiv {\Lambda^2 \over 2M}={8 \over 5}|{\bar \mu}|.
\label{eq.Lam}
\end{eqnarray}
We note that $\omega_c$ is comparable to the binding energy $E_{\rm bind}=2|{\bar \mu}|$ of a molecule in the BEC limit. Thus we conclude that $\omega_c$ is a {\it physical} cutoff determined by the dissociation energy of a molecule in the BEC phase.
\par
To summarize, when we replace $V_{\rm eff}^{(4)}(q,q',K)$ and $V_{\rm eff}^{(3)}(q,K)$ in (\ref{eq.7}) with $V_{\rm eff}^{(4)}(0,0,0)$ and $V_{\rm eff}^{(3)}(0,0)$, respectively, we have to include the {\it physical} momentum cutoff $\Lambda$, as discussed above. We recall that $\Phi_q=\delta\Delta_q/\eta$ describes {\it excitations} from the Bose-condensate. The molecular Bose field involving the condensate part is given by $\phi_q\equiv \Phi_q+\phi_M$. Using $\phi_q$ in (\ref{eq.7}), we obtain
\begin{eqnarray}
S_\Delta
&=&
{1 \over \beta}
\sum_{q}
\phi_q^\dagger
(-i\nu_n+\varepsilon_q^M-\mu_M)
\phi_q
\nonumber
\\
&+&
{1 \over 2\beta^3}
{4\pi(2a_{\rm F}) \over M}
\sum_{q,q',K}^\Lambda\phi_{{K \over 2}+q'}^\dagger\phi_{{K \over 2}-q'}^\dagger\phi_{{K \over 2}-q}\phi_{{K \over 2}+q},
\label{eq.14}
\end{eqnarray}
where $\mu_M\equiv U_Mn_c$ is the molecular chemical potential. We note that $V_{\rm eff}^{(3)}$ has been absorbed into the last term of (\ref{eq.14}). The corresponding Hamiltonian has the form
\begin{eqnarray}
H_{\rm BEC}
&=&
\sum_{\bf q}
\phi_{\bf q}^\dagger
(\varepsilon_{\bf q}^M-\mu_M)
\phi_{\bf q}
\nonumber
\\
&+&
{4\pi(2a_{\rm F}) \over M}
\sum_{q,q',K}^\Lambda\phi_{{K \over 2}+q'}^\dagger\phi_{{K \over 2}-q'}^\dagger\phi_{{K \over 2}-q}\phi_{{K \over 2}+q}.
\label{eq.15}
\end{eqnarray}
This Hamiltonian is also obtained in the normal phase, but now with $\mu_M=4(\mu+\sqrt{|\mu||{\bar \mu}|})$. 
\par
The effective Hamiltonian in (\ref{eq.15}) gives a clear physical picture of the two different molecular scattering lengths found in earlier literature, namely $a_{\rm M}=2a_{\rm F}$\cite{Randeria,Engelbrecht} and $a_{\rm M}=(0.6\sim0.75)a_{\rm F}$\cite{Strinati,Petrov}. Although $U_M=4\pi(2a_{\rm F})/M$ looks like the low-energy expression for $s$-wave interaction, (\ref{eq.15}) shows that $U_M$ is actually a {\it bare} interaction involving a cutoff energy $\omega_c$. In the case of two molecules in a vacuum, to obtain the low-energy interaction $U_M^{2b}\equiv4\pi a_{\rm M}^{2b}/2M$, we have to {\it renormalize} $U_M$ to remove the need for a high-energy cutoff\cite{Pethick}. In the usual manner, this is given by
\begin{eqnarray}
{4\pi a_{\rm M}^{2b} \over M}
&=&
{{4\pi(2a_{\rm F}) \over M}
\over
1+{4\pi (2a_{\rm F}) \over M}
\sum^{\omega_c}{1 \over 2\varepsilon_{\bf q}^M}
}
=
{4\pi \over M}{2a_{\rm F} \over 1+2.28}
\nonumber
\\
&=&
{4\pi (0.61a_{\rm F}) \over M}.
\label{eq.16}
\end{eqnarray}
Here we note that $a_{\rm F}$ in the denominator has been canceled out by the $a_{\rm F}$ involved in $\omega_c$, defined in (\ref{eq.Lam}). The resulting renormalized Hamiltonian has the same form as (\ref{eq.15}), but now with $0.61a_{\rm F}$ in place of $2a_{\rm F}$\cite{noteC}. Eq. (\ref{eq.16}) agrees with the recent four-fermion (two-molecule) analysis, which gives $a_{\rm M}=0.6a_{\rm F}$\cite{Petrov}. Thus, we find that $a_{\rm M}=2a_{\rm F}$ is the bare molecular scattering length before renormalization, while $a_{\rm M}=(0.6\sim0.75)a_{\rm F}$ is the {\it renormalized} two-molecule scattering length in which the high-energy processes up to $\omega_c$ have been incorporated.
\par
The effective action in (\ref{eq.14}) is also useful to calculate the molecular scattering length $a_{\rm M}$ in a {\it many-particle} system. In this case, $a_M$ is affected by many-body effects, as well as by temperature, which are not included in the $T=0$ two-molecule result in (\ref{eq.16}). To include these effects, we apply the RG theory developed for an atomic Bose-condensed gas\cite{Stoof} to deal with (\ref{eq.14}), including the physical cutoff $\Lambda$. In the 1-loop level, the RG equations are given by\cite{Stoof}, for $\mu_M<0$,
\begin{eqnarray}
{d\mu_M \over dl}=2\mu_M-{\Lambda^3 \over \pi^2}U_MN(\varepsilon^M_\Lambda-\mu_M),
\label{eq.17}
\end{eqnarray}
\begin{eqnarray}
{dU_M \over dl}
&=&
-U_M-{\Lambda^3 \over 2\pi^2}U_M^2
\Bigl[
{2N(\varepsilon^M_\Lambda-\mu_M)+1 \over 2(\varepsilon_\Lambda^M-\mu_M)}
\nonumber
\\
&+&
4\beta 
N(\varepsilon^M_\Lambda-\mu_M)
[N(\varepsilon^M_\Lambda-\mu_M)+1]
\Bigr],
\label{eq.18}
\end{eqnarray}
and for $\mu_M\ge 0$\cite{note},
\begin{eqnarray}
{d\mu_M \over dl}
&=&
2\mu_M-{\Lambda^3 \over 2\pi^2}U_M
\Bigl[
{\varepsilon_\Lambda^M+\mu_M \over E_\Lambda^M}
[2N(E^M_\Lambda)+1]-1
\nonumber
\\
&+&
4\beta
N(E^M_\Lambda)[N(E^M_\Lambda)+1]
\Bigr],
\label{eq.19}
\end{eqnarray}
\begin{eqnarray}
{dU_M \over dl}
&=&
-U_M-{\Lambda^3 \over 2\pi^2}U_M^2
\Bigl[
{2N(E^M_\Lambda)+1 \over 2E_\Lambda^M}
\nonumber
\\
&+&
4\beta 
N(E^M_\Lambda)
[N(E^M_\Lambda)+1]
\Bigr].
\label{eq.20}
\end{eqnarray}
Here, $E_\Lambda^M=\sqrt{\varepsilon_\Lambda^M(\varepsilon_\Lambda^M+2\mu_M)}$, and $N(\varepsilon)$ is the Bose distribution function. $\mu_M$, $U_M$, and $T$ involve the trivial scaling as $\mu_M(l)=\mu_M e^{2l}$, $U_M(l)=U_M e^{-l}$, $T(l)=T e^{2l}$. We solve the RG equations together with the equations for the number of molecules $n_{\rm M}=n_{\rm F}/2=n_c+n_n$. 
Here, $n_c=\lim_{l\to\infty}\mu_{\rm M}(l)/U_{\rm M}(l) e^{-3l}$ ($\mu_{\rm M}(l)>0$) is the number of Bose-condensed molecules, and the number of non-condensate molecules $n_n$ is determined by\cite{Stoof},
\begin{eqnarray}
n_n=
{\Lambda^3 \over 2\pi^2}
\int_0^\infty dl N(\varepsilon_\Lambda^M-\mu_M) e^{-3l}~~~(\mu_M<0),
\label{eq.21}
\end{eqnarray}
\begin{eqnarray}
n_n
&=&
{\Lambda^3 \over 2\pi^2}
\int_0^\infty dl 
\Bigl[
{\varepsilon_\Lambda^M+\mu_M \over 2E_\Lambda^M}
[N(\varepsilon_\Lambda^M-\mu_M)
+
1]
-
{1 \over 2}
\Bigr]
e^{-3l}
\nonumber
\\
&{ }&~~~~~~~~~~~~~~~~~~~(\mu_M\ge 0).
\label{eq.22}
\end{eqnarray}
\begin{figure}
\includegraphics[width=9cm,height=6cm]{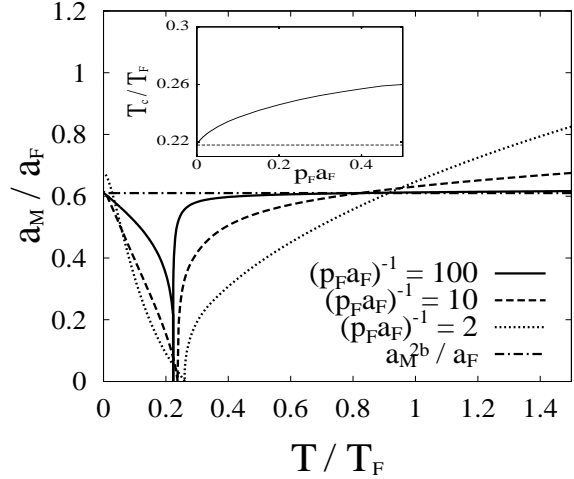}%
\caption{
Many-body molecular scattering length $a_M$ as a function of temperature. $p_{\rm F}$ is the Fermi momentum for a free Fermi gas.
We show results down to $T=0$, although the RG equations (\ref{eq.19}) and (\ref{eq.20}) are not valid for temperatures such that $T/T_{\rm F}\lesssim 0.1(p_{\rm F}a_{\rm F})$.The inset shows $T_{\rm c}$ in the BEC regime determined from the fixed points of the RG equations (\ref{eq.19}) and (\ref{eq.20}). The dashed line shows $T_{\rm c}^{\rm ideal}=0.218T_{\rm F}$. 
\label{fig3}
}
\end{figure}
\par
Figure 3 shows the {\it many-body} molecular scattering length $a_M$ at finite temperatures. The scattering length $a_M$ depends on temperature, and the temperature dependence is stronger as one approaches the crossover regime. At $T_{\rm c}$, $a_{\rm M}$ vanishes. This is expected since if molecules are to be stable at $T_{\rm c}$, the scattering between them must vanish at $T_{\rm c}$. 
We note that a similar $T$-dependent scattering length was also obtained earlier in the study of Bose-condensed gases\cite{Stoof,Griffin}. 
However, in the present situation, one may have such
large values of $a_{\rm F}$ near the resonance that the decrease
in the effective value of $a_{\rm M}$ may have observable
consequences at finite temperatures. In atomic
Bose gases, the small size of the scattering length
meant this region was restricted to temperatures very
close to $T_{\rm c}$ and hence of less interest.
\par
As shown in the inset in Fig.3, the many-body effects enhance $T_{\rm c}$ compared with $T^{\rm ideal}_{\rm c}=0.218T_{\rm F}$ for the transition temperature of a non-interacting Bose gas. In the extreme BEC limit, we find $\Delta T_{\rm c}\equiv T_{\rm c}-T_{\rm c}^{\rm ideal}\propto (p_{\rm F}a_{\rm F}) \propto (n_{\rm M}^{1/3}a_{\rm M})$, consistent with the previous work on the BEC in Bose gases\cite{Stoof,Gruter,Baym,Arnold}. Although $T_{\rm c}$ in the inset initially increases with $p_{\rm F}a_{\rm F}$ from the BEC side, it will eventually decrease in the crossover region, smoothly going into the weak-coupling BCS result\cite{Ohashi}.
\par
Our theory is only valid in so far as the effect of fluctuations involving terms higher order than $n=4$ in (\ref{eq.4}) is small. Comparing the fourth order term with the sixth order term above $T_{\rm c}$, we find that the fourth order term is dominant as long as $(p_{\rm F}a_{\rm F})^{-1}\gesim 1$, i.e., we are not in the unitarity region.
\par
To conclude, we have studied the effective interaction between bound molecules in the BEC regime of a superfluid Fermi gas. We have included the binding energy of a molecule, which naturally leads to a physical cutoff energy in the molecular Bose gas. For the two-molecule case, we showed that this cutoff gives a simple physical explanation of the difference between the two molecular scattering lengths, the bare value $a_{\rm M}=2a_{\rm F}$\cite{Randeria,Engelbrecht} and the renormalized value $a_{\rm M}=(0.6\sim 0.75)a_{\rm F}$\cite{Strinati,Petrov}. 
We have also shown that the {\it many-body} scattering length deviates considerably from the two-body scattering length as the temperature approaches $T_{\rm c}$ if we are close to the unitarity limit. By employing the RG technique, the present paper includes fluctuations past the mean-field BCS theory in a more sophisticated way than the original crossover theory developed by Nozi\`eres and Schmitt-Rink\cite{Nozieres}.
\par
I would like to thank Allan Griffin for stimulating discussions and comments, as well as a critical reading of this paper. This work was supported by funds from Ministry of Education of Japan and from NSERC Canada.
%

%

\begin{thebibliography}{99}
\bibitem{notez} Including the $K$-dependence of $V_{\rm eff}^{(4)}$, we obtain another cutoff for the center-of-mass momentum as $\Lambda_K\equiv\sqrt{32/3}/a_{\rm F}$, which corresponds to the energy of an incident molecule given by $\varepsilon^M_{\Lambda_K/2}=(4/3)|\mu|$. We note that $\Lambda_K$ is not important in determining the low-energy value of $a_M$.
\bibitem{noteC} The numerical factor $0.61$ in (\ref{eq.16}) weakly depends on the way of introducing the cutoff in the present analysis. When we use the smooth cutoff $F(q)F(q')$ for the molecular interaction in (\ref{eq.15}), we obtain $a_{\rm B}^{2b}=0.71a_{\rm F}$.
\bibitem{note}
The RG equations in the 1-loop level has an infrared divergence below $T_{\rm c}$\cite{Stoof}. We have eliminated this divergence, using the prescription\cite{Stoof} that one replaces the coefficients in the Bogoliubov transformation as $(u_{\bf p},v_{\bf p})=(1,0)$. This method is valid when we can approximate $E_{\bf p}=\sqrt{\varepsilon_{\bf p}^M(\varepsilon_{\bf p}^M+2\mu_M)}\simeq\varepsilon_{\bf p}^M+\mu_M$. Since the temperature works as a low energy cutoff, this condition can be written as $T\gesim\mu_M$, or $T/T_{\rm F}\gesim 0.1p_{\rm F}a_{\rm F}$ (assuming the $T=0$ result $a_{\rm M}=0.6a_{\rm F}$). Because $(p_{\rm F}a_{\rm F})^{-1}\gg 1$ in the BEC region, this condition is largely satisfied, except at very low temperatures.
\end{thebibliography}
\end{document}